\documentclass[a4paper,
               keeplastbox,   
               ]{jacow}
\usepackage{pdfpages,multirow,ragged2e,textgreek,mathtools} %
\makeatletter%
	\ifboolexpr{bool{xetex}}
	 {\renewcommand{\Gin@extensions}{.pdf,%
	                    .png,.jpg,.bmp,.pict,.tif,.psd,.mac,.sga,.tga,.gif,%
	                    .eps,.ps,%
	                    }}{}
\makeatother

\ifboolexpr{bool{xetex} or bool{luatex}} 
 {}                                      
 {\usepackage[utf8]{inputenc}}           

\usepackage[USenglish]{babel}

\ifboolexpr{bool{jacowbiblatex}}%
 {%
  \addbibresource{jacow-test.bib}
  \addbibresource{biblatex-examples.bib}
 }{}
\DeclareSIUnit{\sample}{S}

\begin{document}

\title{A NEW LONGITUDINAL DIAGNOSTICS SYSTEM FOR CERN\textquotesingle S ANTIPROTON MACHINES}

\author{M. E. Angoletta\thanks{maria.elena.angoletta@cern.ch}, D. Barrientos, B. Bielawski, M. Jaussi, \\M. Niccolini, A. Rey, M. Soderen, M. Suminski, CERN, Geneva, Switzerland \\
		J. Molendijk, retired staff, CERN, Geneva, Switzerland \\
		V. R. Myklebust, S. Novel Gonzalez, former fellows, CERN, Geneva, Switzerland }
	
\maketitle

\begin{abstract}
A new powerful longitudinal diagnostics is being developed for the two CERN’s antiproton machines, the Antiproton Decelerator (AD) and the Extra Low ENergy Antiproton (ELENA) ring. The longitudinal diagnostics receives data from the Low-Level Radio Frequency (LLRF) system via optical fiber. Real time processing of gigabit data-streams is enabled by its computational resources and real-time operating system. Measurement of intensity, bunch length and turn-by-turn bunch profiles are available for bunched beams. Intensity, momentum spread and average frequency will soon be measured for debunched beams. The system will provide essential input to setup and monitor RF and cooling systems. It will also be used by operators to monitor each machine’s performances and the overall efficiency of the antiproton chain. This paper shows the data acquisition/processing mechanism and preliminary bunched beam results for AD. Next steps and plans for exporting the system to other CERN machines equipped with the same LLRF system are also mentioned.
\end{abstract}

\section{INTRODUCTION}
Traditional beam transformers don’t work at the low intensities of up to $4e7$ antiprotons present in CERN’s Antiproton Decelerator (AD) and Extra Low Energy Ion Ring (ELENA). Alternative ways of measuring beam intensity and other characteristics must be utilized. A dedicated system, processing bunched and debunched beam data, was available in AD from its first run in 2000 to 2018 \cite{AD_INT}. ELENA's commissioning \cite{ELENA_COMM} and the AD consolidation \cite{AD_COMM} then required a new intensity measurement system for both machines. 

Intensity and bunch measurements from frequency-domain data are available as part of the Low-Level RF (LLRF) systems for ELENA and AD. A cryogenic current comparator \cite{CCC} system, installed in the AD since 2018, measures bunched and debunched beam intensity but suffers from reliability issues. This paper describes a novel system to measure intensity, bunch length and turn-by-turn bunch profiles from time-domain data. Intensity, momentum spread and average frequency will soon be measured for debunched beams, too. Figure 1 shows the AD cycle as an example of interleaved bunched and debunched beam segments. The red boxes are time windows when the beam is bunched, referred to as "RF segments". The number written in each box is the RF harmonic number for that segment. 
\begin{figure}[!htb]
   \centering
   \includegraphics*[width=.9\columnwidth]{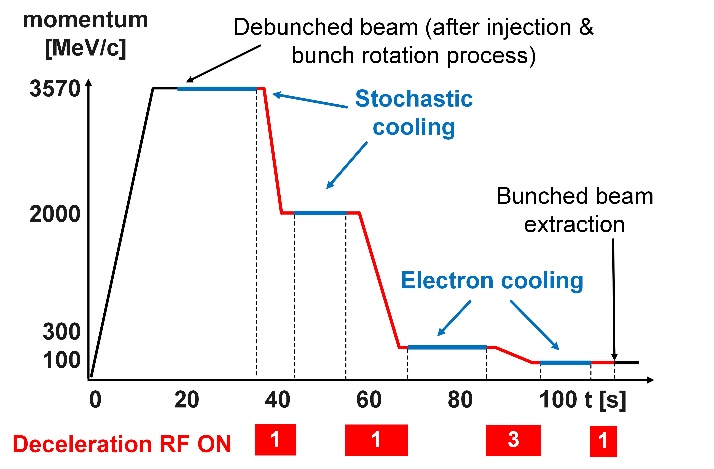}
   \caption{AD production cycle.}
   \label{fig:figure1}
\end{figure}

\section{System layout}
\subsection{AD System Layout}
\begin{figure*}[!tbh]
    \centering
    \includegraphics*[width=0.9\textwidth]{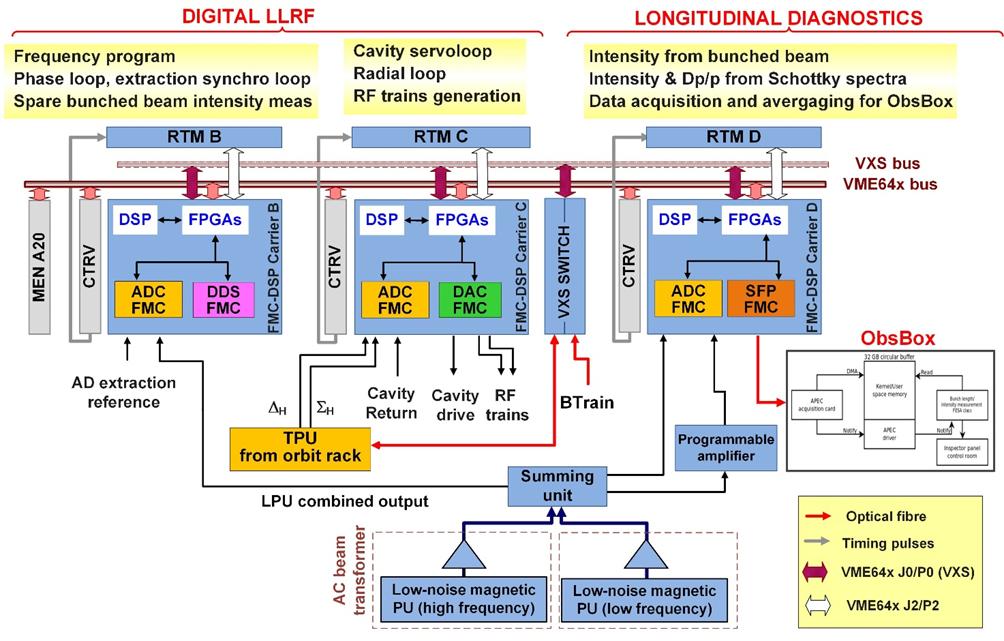}

    \caption{Longitudinal diagnostics as part of the AD LLRF system. Keys: FMC – FPGA Mezzanine Card, DDS –Direct Digital Synthesiser, ADC – Analogue-to-Digital Converter, DAC – Digital-to-Analogue Converter, SFP – Small Form-factor Pluggable Transceiver, LPU/TPU – Longitudinal/Transverse Pick-Up, CTRV – Timing Receiver Module, MEN A20 – Master VME board, RTM – Rear Transition Module, ObsBox – custom processing subsystem. }
    \label{fig:figure2}
\end{figure*}

Figure~\ref{fig:figure2} shows the AD LLRF system as an example of the longitudinal diagnostics integration within the LLRF system for both AD and ELENA. 
The LLRF system is clocked with a fixed, \SI{122.7}{\mega\hertz} clock. The FMC-DSP carrier D is in charge of longitudinal diagnostics and hosts a 4-channel ADC and a SFP FMC daughter-cards. In the AD only the output from a longitudinal magnetic pick-up is connected to the ADC FMC, amplified via  a programmable amplifier for the future debunched beam analysis. The FMC-DSP carrier D is responsible for sampling the analogue signals, pre-processing them and sending them to the data processing system, referred to as ObsBox. Additional information such as beam status (bunched vs debunched), beam harmonic and revolution frequency is also sent. In ELENA the sum output from the Transverse pick-up, used by the LLRF radial loop, is sent to the ObsBox system, too, as it currently shows less noise that the ELENA longitudinal magnetic pick-up. 

\begin{figure}[!htb]
   \centering
   \includegraphics*[width=\columnwidth]{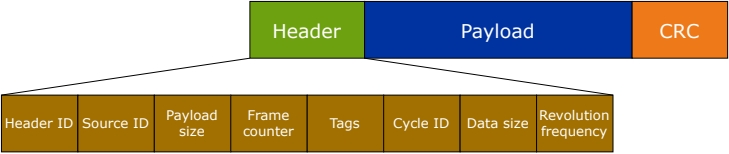}
   \caption{Frame composition.}
   \label{fig:figure3}
\end{figure}

An FPGA in the carrier board receives the digitized data and low-pass filters them to a fixed sampling rate of \SI{61.357}{\mega\sample\per\second}. These digital data are continuously stored using a double-buffer scheme on the device internal memory resources. The FMC-DSP carrier B generates and distributes a new revolution frequency value every \SI{12.5}{\micro\second}. 
A new frame is created and the memory buffer is swapped when a new frequency value is received. Frames are transmitted over optical fibers to the ObsBox subsystem  via high-speed digital links. The receiver side is locked to the data stream via standard comma characters. Frames are then decoded and processed individually for each channel by the ObsBox.
\newpage
Figure~\ref{fig:figure3} shows the main components of a data frame. They include: a) a header with a fixed size of eight 32-bit words; b) a payload slot of variable size composed by 16-bit signed samples recorded since the last revolution frequency value was received; c) a 32-bits Cyclic Redundancy Check (CRC) value computed for the whole frame. The organisation of words in the frame header is summarised in Table 1.
\begin{table}[!hbt]
	\setlength\tabcolsep{3.5pt}
	\centering
	\caption{Organisation of words in frame header}
	\label{tab:styles}
\begin{tabular}{||c | c||} 
 \hline
 Word & Description  \\ 
 \hline\hline
 Header ID & Fixed constant \\ 
 \hline
 Source ID & Channel identifier \\
 \hline
 Payload size & Number of bytes in payload \\
 \hline
 Frame counter & Incremented each frame \\
 \hline
 Tags & Beam state, RF harmonic number \\ 
 \hline
  Cycle ID & Machine cycle identifier \\
 \hline
  Data size & Data word size in payload slot \\
 \hline
   Revolution frequency & Current revolution frequency \\
 \hline
\end{tabular}
\end{table}

\subsection{ObsBox hardware layout}
The ObsBox system was originally developed for analysis of the transverse plane in CERN’s Large Hadron Collider \cite{OBSBOX_LHC}. It has been constantly evolving and is now being adopted in most of CERN' accelerators.

The system consists of commercial hardware coupled with custom firmware and driver which allows for fast transfer from the serial link to the host. Figure 4 provides a simplified system overview. It can receive data at a rate of up to 10 Gbps, in AD and ELENA the current configuration is two channels with a rate-rate of \SI{2}{\giga\bit\per\second}.

This data is transferred by the hardware to a circular buffer in the host's RAM \cite{OBSBOX_2019}. The driver is notified as soon as a transfer is completed. Multiple applications can concurrently access the circular buffer, polling the driver for new data. One of these applications is a FESA \cite{FESA} class that receives the un-decimated data-stream from the ADCs, sampling the signal from the longitudinal pickup. This class analyses the stream to calculate intensity and bunch profile which can be presented to the operators of the machines. 

\begin{figure}[!htb]
   \centering
   \includegraphics*[width=\columnwidth]{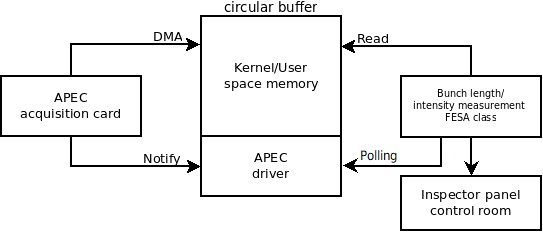}
   \caption{ObsBox system overview.}
   \label{fig:figure4}
\end{figure}

\section{Bunched beam data Processing}
\subsection{Overview}
This section outlines the Obsbox subsystem that produces typical measured characteristics for a beam, such as intensity, beam length, and peak value.
The ADC data stream and additional information, such as RF harmonic and revolution frequency, are treated by the FESA application through the pipeline of elementary processing blocks shown in Figure 5. More details on the processing blocks are provided below.
\begin{figure}[!htb]
   \centering
   \includegraphics*[width=\columnwidth]{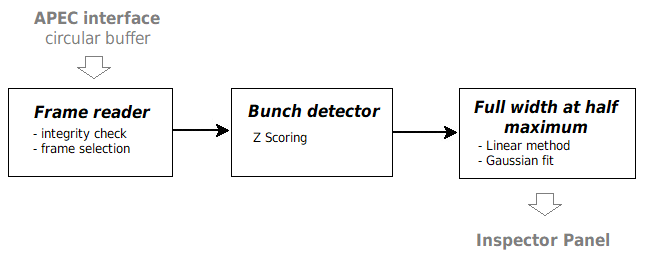}
   \caption{Bunch length/ intensity measurement FESA class block scheme }
   \label{fig:figure5}
\end{figure}

\subsection{Frame reading and bunch detection}
The frame reader accesses the circular buffer through the APEC interface and builds batches of consecutive samples thanks to the beam state tag in the frames' header. The first step in the algorithm is to detect a bunch in the stream of samples. This is achieved by a peak detector using the smoothed Z-score algorithm \cite{Z_SCORE} that gives to each sample in a time-series a z-score:
\begin{equation} 
\label{eq:z-score}
 z=\frac{x-\hat{x}}{S}
\end{equation} 
where $\hat{x}$ is the mean of previous samples in the time-series and $S$ is the standard deviation of the sample. If $z$ is greater than a certain threshold then the time-series contains a bunch. The stream is chunked with a fixed size around the threshold-crossing point. For each chunk \textit{baseline} and \textit{peak value} are computed to define bunch rising and falling edges at full width at half maximum (FWHM). The baseline is found as the lowest average of consecutive $n$ samples in the chunk, where $n$ is selectable, taken before the peak occurrence to avoid bias due to undershooting.
Chunks are then processed to extract the characteristics of the contained bunch.

\subsection{Linear Method}
Once bunch start and end are defined at FWHM, the \textit{linear bunch length} is obtained as the number of samples within the bunch edges multiplied by the ADC sampling interval. To obtain the intensity via the linear method, the Riemann integral of the samples is calculated, after subtracting the baseline from the samples amplitude, as shown in Eq.~\ref{eq:intensity}.
\begin{equation} 
\label{eq:intensity}
 N_{P,lin}=k \cdot \sum_{\text{bunch start}}^{\text{bunch end}}V_{sample}
\end{equation} 
where k is a multiplication factor including the amplitude of each sample, the gain of the programmable amplifier, the ADC sampling interval, the harmonic number and a fudge factor to calibrate the measurement. A \textbeta -normalization factor, proportional to the revolution frequency, is introduced when the input originates from a Transverse pick-up.
\subsection{Gaussian fit}
Assuming a Gaussian shape of the bunch and using customised algorithms already devised in the RF group \cite{BQM}, samples are fitted to a Gaussian function:

\begin{equation} 
\label{eq:gaussian}
 X[n]=a\cdot e^{-\dfrac{(n-b)^2 }{2c^2} }
\end{equation} 
Which can be rewritten to:
\begin{equation} 
\label{eq:gaussianlog}
 ln(X[n])=ln(a) - \dfrac{(n-b)^2 }{2c^2} 
\end{equation} 
\begin{equation} 
\label{eq:gaussianlog}
 ln(X[n])=ln(a) - \dfrac{b^2 }{2c^2} +\dfrac{bn}{c^2}-\dfrac{n^2}{2c^2}
\end{equation} 
with:

\bigskip\noindent
$\hfill A=ln(a) -\dfrac{b^2}{2c^2} \hfill B=\dfrac{b}{c^2} \hfill C=-\dfrac{1}{2c^2}\hfill$
\bigskip\noindent

the resulting equation is:
\begin{equation} 
\label{eq:gaussianfinal}
ln(X[n])=A+B\cdot n + C \cdot n^2 
\end{equation} 
Using quadratic regression \cite{REGRESSION} the samples can be fitted according to Eq.~\ref{eq:gaussianfinal} to find the coefficients $A$, $B$, and $C$, allowing to write the Gaussian function parameters $a$, $b$, $c$ of Eq.~\ref{eq:gaussian} as:

\bigskip\noindent
$\hfill a=e^{A-\dfrac{B^2}{4C}} \hfill b=-\dfrac{B}{2C} \hfill c=\sqrt{-\dfrac{1}{2 C}}$
\bigskip\noindent

The intensity is proportional to the integral of the fitted Gaussian function in Eq.~\ref{eq:gaussian} and this is done in the interval of 4 sigma which gives:
\begin{equation} 
\label{eq:gaussianInt}
\begin{split}
 & \int_{b-2c}^{b+2c} a\cdot e^{-\dfrac{(n-b)^2 }{2c^2} }=\sqrt{2 \pi} \cdot a \cdot c\cdot erf(\sqrt{2})\\
 &=\sqrt{2\pi}e^{(A-\dfrac{B^2}{4C})} \cdot \sqrt{-\dfrac{1}{2C}}\cdot erf(\sqrt{2})
\end{split}
\end{equation} 
where $erf$ is the Gauss error function. When the input signal come from a Transverse pick-up, a \textbeta -normalization is needed.

Gaussian bunch length is computed proportionally to the FWHM as in Eq.~\ref{eq:fhwm}. The standard deviation ($\sigma$) in Eq.~\ref{eq:gaussian} is $c$.
\begin{equation} 
\label{eq:fhwm}
\begin{split}
FWHM = &2\sqrt{2ln(2)}\sigma \\
 = &2\sqrt{2ln(2)}\sqrt{-\dfrac{1}{2 C}}
\end{split}
\end{equation} 

\section{Bunched beam results}
This paragraph describes applications presenting the measured bunched beam parameters to users. Their aim is to allow machine operators to monitor and document on the elogbook the performance of the machine. They also help RF experts with their systems setup. The applications were developed with Inspector \cite{INSPECTOR}, a zero-code IDE for Control Systems GUI development devised by CERN.

\subsection{Cycle Monitoring Display}
The Cycle Monitoring Display shows revolution frequency, beam intensity and bunch length  as a function of the time in the cycle, referred to as \textit{ctime}. Data is refreshed at the end of each cycle and the parameters are shown for a subset of the detected bunches. The subset can be customized on-the-fly by selecting the time interval between extracted bunches (25 \textmugreek s – 250 ms) and averaging the characteristics over a selectable number of consecutive bunches (1-500).
Figure 6 shows the cycle monitoring panel for an AD cycle; at the top a summary of the intensities and length in the various RF segments is shown as well as the plot of the revolution frequency from the LLRF system. Beam intensity and bunch length are also displayed, plotting the measurements obtained via a Gaussian fit in blue and the linear method in red.
\begin{figure}[!htb]
   \centering
   \includegraphics*[width=\columnwidth]{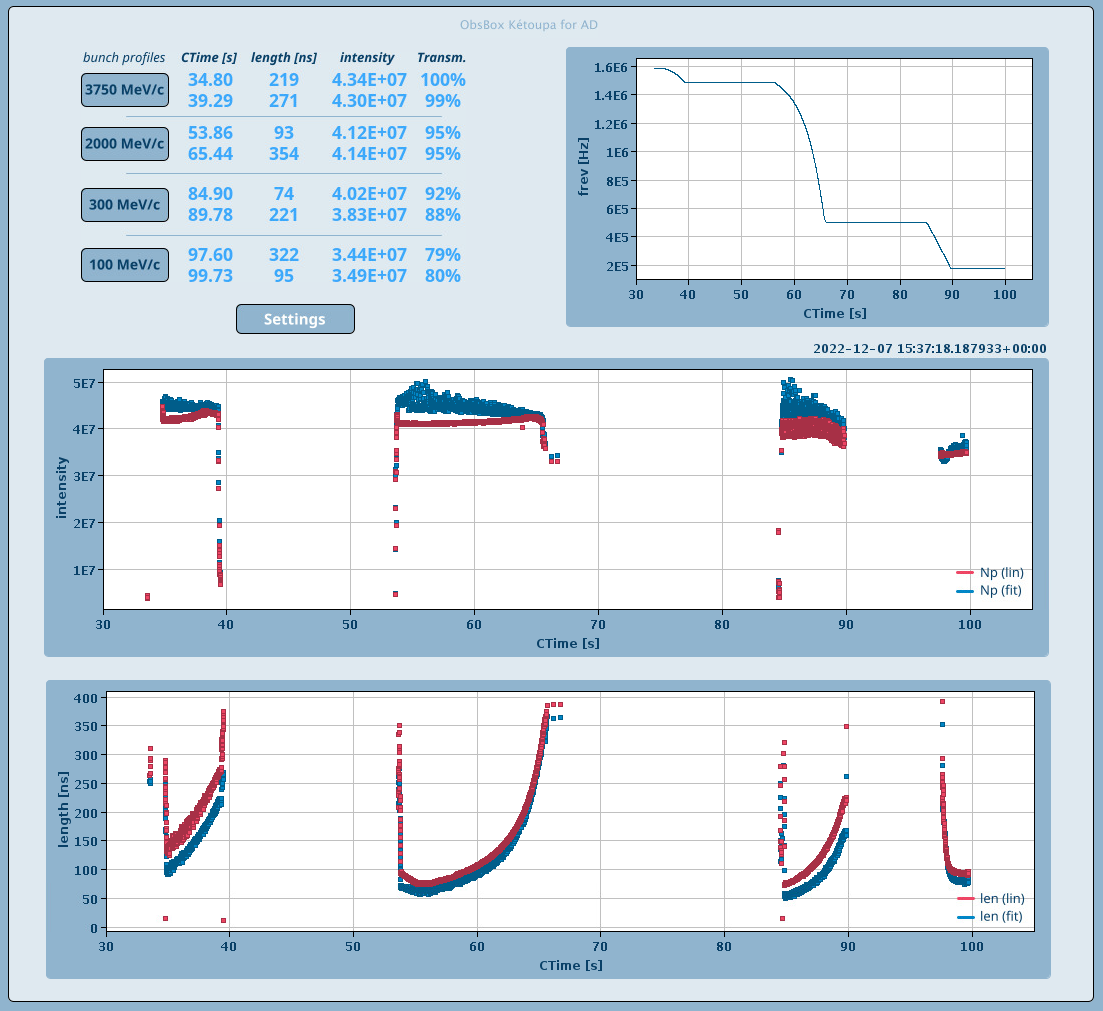}
   \caption{From top to bottom: plot of revolution frequency, of beam intensity and of bunch length for an AD cycle. Measurements obtained via a Gaussian fit are shown in blue and those obtained through a linear methd are in red.  }
   \label{fig:figure6}
\end{figure}
The bunch length is very high at the beginning of each segment as it corresponds to the start of the bunching process. A difference between the two intensity calculation methods is also visible: in particular, Figure 7 zooms into the last RF segment, evidencing how the Gaussian method slight diverges from the linear one. The reasons for this are currently being evaluated both in the quadratic regression method that estimates the coefficients in Eq.~\ref{eq:gaussianfinal} and in the selection of samples used in the algorithm.
\begin{figure}[!htb]
   \centering
   \includegraphics*[width=\columnwidth]{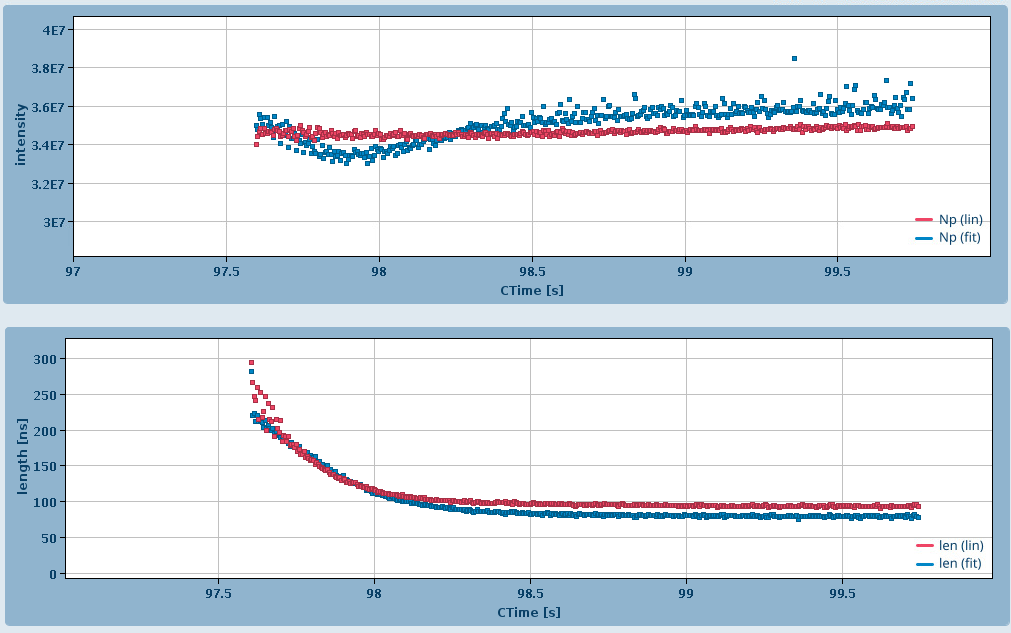}
   \caption{Bunched beam intensity and bunch length on the AD extraction plateau, comparing the Gaussian (blue trace) and linear (red trace) methods for calculating beam intensity and beam length. }
   \label{fig:figure7}
\end{figure}
At the beginning of this segment is possible to observe a distorted bunch shape due to the bunched beam cooling carried out on the extraction plateau to reduce the bunch length at a constant revolution frequency. The interaction between electron cooling and beam capture causes a distorted bunch shape at the beginning of the RF segment, as shown in Figure 8.
\begin{figure}[!htb]
   \centering
   \includegraphics*[width=\columnwidth]{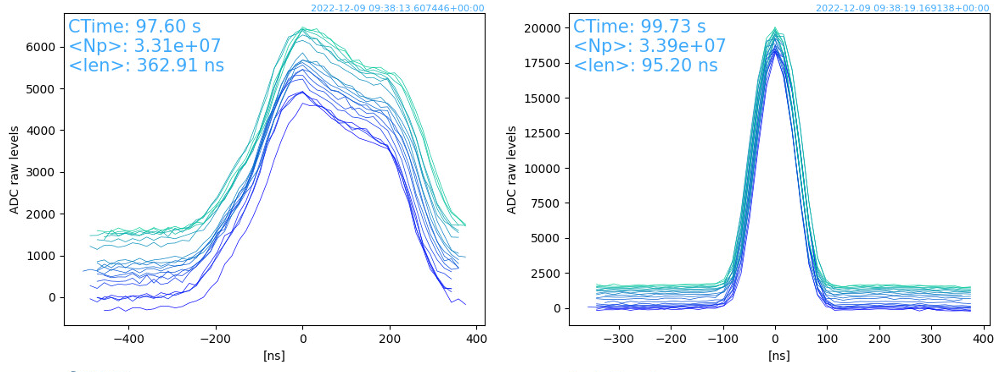}
   \caption{Bunch profiles at beginning and end of the extraction plateau, the first 20 bunches at the indicated \textit{ctime} are shown. The bunch shape at capture is non-gaussian owing to the concurrent electron cooling.}
   \label{fig:figure8}.
\end{figure}

\subsection{Machine efficiency display}
The table in the top left corner of Figure 6 provides intensity values acquired at precise points in the cycle, so that machine operators can evaluate the deceleration efficiency. Intensity and bunch length measured with linear method at capture and at the end of each decelerating ramp are shown in each line. In the third RF segment (300 MeV/c) the RF harmonic is set to three, there are three bunches in the machine, so the bunch lengths are noticeably shorter than in the other segments. The values in the last row refer to the bunch intensity and length immediately before extraction. The right-most column calculates the percentage of beam remaining in the machine with respect to the first bunched beam intensity measured, thus indicating the deceleration efficiency. Measurements are averaged over a selectable number of consecutive bunches extracted at the specified \textit{ctime}. Variations of baseline and bunch shapes result in the displayed fluctuation in intensity measurement within the same RF segment: a more robust measurement method is currently under investigation.
In the 2023 run the values for AD and ELENA will be combined into a single display, allowing an evaluation of the deceleration efficiency for the whole antiproton chain.

\subsection{Waterfall Plot}
For each RF segment, the profiles and characteristics such as intensity and length of bunches at capture and at the end of the decelerating ramp can be observed. The FESA class also allows to extract such data at multiple, custom user-selected times expressed in \textit{ctime}. A typical usage scenario is to visualize the evolution of the bunch shape during the RF segment, as shown in Figure 9. The bunches in the plot are centered to the peak value.
\begin{figure}[!htb]
   \centering
   \includegraphics*[width=\columnwidth]{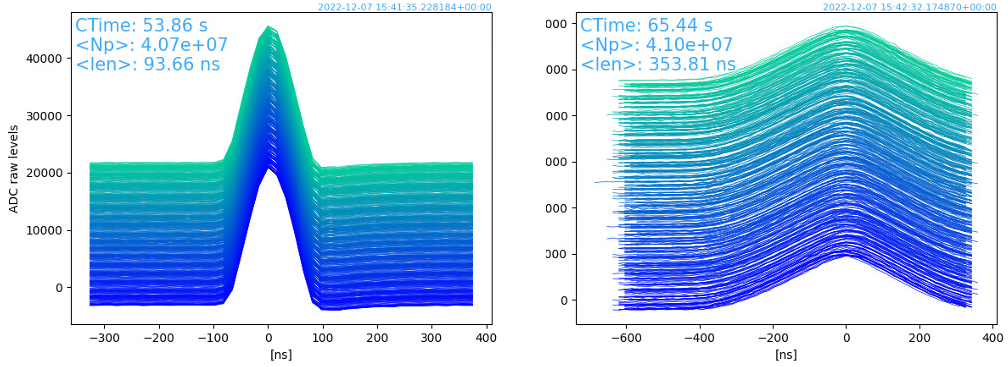}
   \caption{Waterfall plot at the beginning and at the end of the second AD ramp. 500 bunches, recorded turn by turn, are displayed in a mountain range view.}
   \label{fig:figure9}
\end{figure}

\section{NEXT STEPS AND FUTURE WORK}
Cross-calibration between different measurements, in AD and in ELENA, will be done to make sure they are all coherent. 
Software rules will be deployed to automatically adapt the acquisition time for the cycle efficiency display to changes in the AD/ELENA cycle lengths. Debunched beam data processing will also be carried out, to measure  beam intensity, momentum spread and average frequency via Schottky analysis. These values will help the cooling experts with the setup of their systems. 

Finally, the bunched-beam processing will be exported to CERN's Low Energy Ion Ring and PS Booster accelerator, equipped with the same LLRF system as AD and ELENA \cite{REF_LLRF}. Their ObsBox systems will not focus on intensity measurements, as standard intensity diagnostics are already available in these machines. They will instead provide longitudinal beam measurements to better monitor the performance of the LLRF systems. For instance bunch length measurements during a cycle will allow monitoring the effectiveness of the longitudinal beam blowup. They will also provide diagnostics information useful for setting up the LLRF systems. In the longer term, these new and powerful longitudinal diagnostics will allow implementing additional alarms/warnings and slow controls of the LLRF  parameters, thus contributing to an automatic system optimization and reduction of the operational workload.
\\

%
%
\ifboolexpr{bool{jacowbiblatex}}%
	{\printbibliography}%
	{%

} 
%
%


\end{document}